\documentclass[12pt]{iopart}
\usepackage{iopams}
\usepackage{setstack}  

\usepackage{graphicx}
\usepackage{dsfont}
\usepackage[colorlinks]{hyperref}
\usepackage[T1]{fontenc}
\usepackage[utf8]{inputenc}
\usepackage{lmodern}
\usepackage{braket}
\usepackage[caption=false]{subfig}

\begin{document}

\title[]{Solving the scattering of $N$ photons on a two-level atom without computation}

\author{Alexandre Roulet$^1$ and Valerio Scarani$^{1,2}$}
\address{$^1$ Centre for Quantum Technologies, National University of Singapore, 3 Science Drive 2, Singapore 117543, Singapore}
\address{$^2$ Department of Physics, National University of Singapore, 2 Science Drive 3, Singapore 117542, Singapore}
\vspace{10pt}
\begin{indented}
\item[]\today
\end{indented}

\begin{abstract}
We propose a novel approach for solving the scattering of light onto a two-level atom coupled to a one-dimensional waveguide. We first express the physical quantity of interest in terms of Feynman diagrams and treat the atom as a non-saturable linear beamsplitter. By using the atomic response to our advantage, a relevant substitution is then made that captures the nonlinearity of the atom, and the final result is obtained in terms of simple integrals over the initial incoming wavepackets. The procedure is not limited to post-scattering quantities and allows for instance to derive the atomic excitation during the scattering event.
\end{abstract}






\section{Introduction}
Solving analytically the scattering of an arbitrary state of light onto a quantum emitter is a long-standing problem in Quantum Optics. Over the past decade, it has become more and more relevant in the context of \emph{waveguide QED} where the light field is strongly confined along a one-dimensional (1D) waveguide, effectively enhancing the light-matter interaction. Rapid experimental progress is being made in this field, with unprecedented coupling efficiency for cesium atoms trapped near a 1D alligator photonic crystal waveguide~\cite{Goban2015}, and almost perfect coupling in superconducting-qubit~\cite{Hoi2013} and quantum-dot~\cite{Javadi2015} based architectures. Such integrable platforms are especially promising for using the atom as a mediator of photon-photon interaction~\cite{Chang2014}, inducing non-trivial correlations at the level of a few photons~\cite{Lodahl2015}, and for testing proposals such as quantum networks~\cite{Kimble2008,Komar2014}.

From the theory point of view, the most elementary system of waveguide QED consists of a single two-level atom coupled to a 1D waveguide. This system has stimulated a lot of research aiming to characterize how would some specific states of light be scattered in an experiment. Of the most notable is the seminal work~\cite{Domokos2002} by P.~Domokos and co-authors who tackled the scattering of a singe-photon pulse as well as coherent states. J.~T.~Shen and S.~Fan later introduced a powerful framework based on the Lippmann-Schwinger (L-S) equation for solving the transport of a single photon~\cite{Shen2005} and the predicted extinction at resonance has been observed experimentally with low-power coherent states~\cite{Hoi2013,Javadi2015}. While the transport of two photons has been successfully addressed immediately after~\cite{Shen2007}, it was not until last year that several theoretical proposals have been put forward to deal in a systematic way with the scattering of $N$ arbitrary photons. Among these are versatile approaches that leave aside the nature of the scatterer and could in principle be applied to a wide variety of systems~\cite{Xu2015,Shi2015}. In the specific case of a two-level atom, an important step has been achieved in~\cite{Shen2015} where the authors extended the L-S framework to the scattering of $N$ photons. Of importance are also the results of M.~Pletyukhov and V.~Gritsev which are derived for a chiral waveguide using an operator formulation of L-S~\cite{Pletyukhov2012}.

The present work is motivated by the fact that the two-level atom is one of the simplest nonlinear scatterer one could think of. As such, there should be a way of describing the scattering event -- including the dynamics -- with few computations. Moreover, this becomes more than a theoretical challenge as current experiments are about to enter the realm of reliably generating arbitrary photonic states~\cite{Gonzalez2015,Gonzalez2016}. Here we propose a systematic method, which for the first time not only explains how the atom induces correlations between the scattered photons, but uses this very knowledge to greatly simplify the theoretical description. The key difference with recent efforts is that the end-user does not need to \emph{compute} the system dynamics and \emph{recalculate} the response function for each arbitrary number of incoming photons $N$. Instead, our main result is the operational translation of the well-known statement that \emph{a two-level atom can only absorb or emit at most one photon at a given time}.

\begin{figure}
\includegraphics[width=0.6\textwidth]{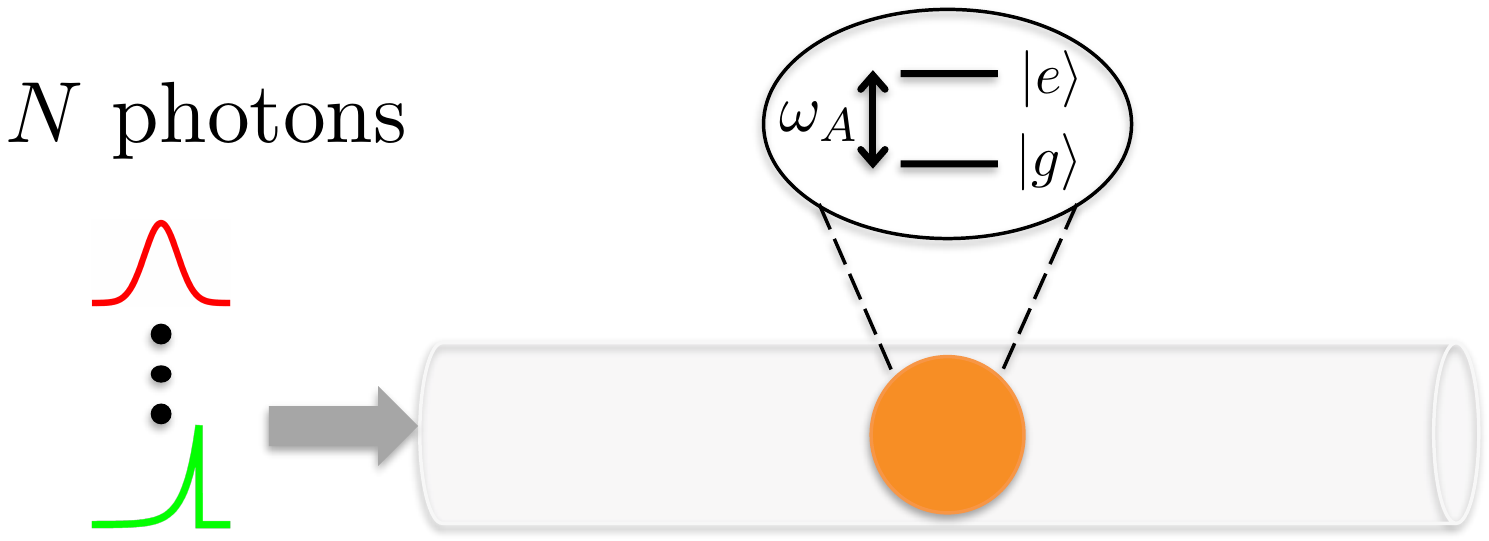}\centering
\caption{\label{fig:schema}Scattering of $N$ photons incoming from the left (\textit{i.e.} only $\xi_N(\tau_1,\ldots,\tau_N)$ contributes in~\eref{eq:lightIn}) onto a two-level atom. The photons do not need to be identical and can have arbitrary profiles, including correlations prior to the scattering event.}
\end{figure}

\section{Model}\label{sec:model}
We consider the system illustrated in~\fref{fig:schema}, consisting of a 1D waveguide strongly coupled to a two-level atom with transition frequency $\omega_A$ between ground $\ket{g}$ and excited $\ket{e}$ states much larger than the cutoff frequency of the waveguide. The dipole Hamiltonian describing the interaction between the atom and propagating photons, under rotating wave approximation, is then given by~\cite{Domokos2002}
\begin{equation}\label{eq:Hamiltonian}
	\hat{H}_{dipole}=-i\hbar\int_{0}^\infty\!\rmd \omega\, g_\omega \Big[\ket{e}\bra{g} \left(\hat{a}^\leftharpoonup_\omega +\hat{a}^\rightharpoonup_\omega \right)-\mathrm{H.c.} \Big] ,
\end{equation}
where $g_\omega$ is the coupling constant and $\hat{a}^\leftharpoonup_\omega$ ($\hat{a}^\rightharpoonup_\omega$) is the annihilation operator of the left- (right-) propagating photon mode at frequency $\omega$. In the following we will work with the Weisskopf-Wigner approximation, where the decay rate of the atom to the waveguide reads $\gamma\equiv 2\pi g^2_{\omega_A}$ and the lower bound of the frequency integration can be extended to $-\infty$~\cite{Scully1997}.

A key step to reveal the physics of the atom's impact on the scattered light is to work in the time domain \cite{Loudon1990}. We will use the following convention for the Fourier transform
\begin{equation}
	\hat{a}_\tau=(2\pi)^{-1/2}\int_{-\infty}^\infty \!\rmd \omega\, \hat{a}_\omega\, \rme^{-\rmi(\omega-\omega_A)\tau},
\end{equation}
where $\omega_A$ is chosen as a reference frequency and $\tau$ is the time distance from the wavefront. The wavefront is associated to the first wavepacket that reaches the atom and sets the start of the scattering event. It therefore coincidences with the atom's position at initial time $t=0$. With this knowledge, we can now express the incoming state for a light field composed of $N$ photons in the time domain as
\begin{equation}\label{eq:lightIn}
	\ket{\psi_N} = \int_{0}^\infty\!\rmd \tau_1\,\cdots\int_{0}^\infty\!\rmd \tau_N\,\sum^{N}_{n=0} \frac{\xi_n(\tau_1,\ldots,\tau_N)}{\sqrt{n!\,(N-n)!}}\prod^{n}_{i=1}\hat{a}^{\rightharpoonup\,\dagger}_{\tau_i}\prod^{N}_{j=n+1}\hat{a}^{\leftharpoonup\,\dagger}_{\tau_j} \ket{\mathrm{vac}} ,
\end{equation}
where $\ket{\mathrm{vac}}$ is the vacuum state of the light field and $\xi_n(\tau_1,\ldots,\tau_N)$ is the normalized wavepacket associated with $n$ photons traveling to the right and the rest traveling to the left. Note that this form allows for initial correlations between the photons and satisfies the bosonic exchange symmetry.

Motivated by the high coupling efficiencies recently achieved experimentally \cite{Hoi2013,Javadi2015}, we will focus on the ideal case where spontaneous emission of the atom to the environment is negligible. Our results are then obtained by deriving the Heisenberg equations for the atom and field operators, which read in the interaction picture (a detailed derivation is given in Appendix A)
\numparts
\begin{eqnarray}\label{eq:sigmazDot}
	\frac{\mathrm{d}}{\mathrm{d}t}\hat{\sigma}_z &=& -2(\mathds{1}+\hat{\sigma}_z)-2(\hat{\sigma}_+\,\hat{d}_\mathrm{in} + \hat{d}_\mathrm{in}^\dagger\, \hat{\sigma}_- ) , \\\label{eq:aDot}
	\frac{\mathrm{d}}{\mathrm{d}t}\hat{a}^\rightharpoonup_\tau&=&\delta(t-\tau)\, \hat{\sigma}_- =\frac{\mathrm{d}}{\mathrm{d}t}\hat{a}^\leftharpoonup_\tau , \\\label{eq:sigma-Dot}
	\frac{\mathrm{d}}{\mathrm{d}t}\hat{\sigma}_- &=& -\hat{\sigma}_- +\hat{\sigma}_z\, \hat{d}_\mathrm{in} ,
\end{eqnarray}
\endnumparts
where time is normalized in units of the atomic lifetime $\gamma^{-1}$ and $\hat{d}_\mathrm{in}(t)= \hat{a}^\leftharpoonup_t(0) +\hat{a}^\rightharpoonup_t(0)$ takes out a photon from the initial state at a distance $t$ from the wavefront. Note that the Pauli matrix and the lowering operator are respectively defined as $\hat{\sigma}_z=\ket{e}\bra{e}-\ket{g}\bra{g}$ and $\hat{\sigma}_-=\ket{g}\bra{e}$. Of particular interest is the solution to the last equation~\eref{eq:sigma-Dot}
\begin{equation}\label{eq:sigma-}
	\hat{\sigma}_-(t)=\rme^{-t}\,\hat{\sigma}_-(0)\,+ \int_{0}^t\!\rmd t'\, \rme^{-(t-t')}\, \hat{\sigma}_z(t')\, \hat{d}_\mathrm{in}(t') .
\end{equation}
Here the first term corresponds to the relaxation of an initially excited atom while the last term is less straightforward to interpret. Indeed, if the atom was to respond like a linear beamsplitter, \textit{i.e.} without being saturated and dealing with each photon as if the others were absent, this last term would effectively have the form
\begin{equation}\label{eq:sigma-Lin}
	\hat{\sigma}^{\mathrm{lin}}_-(t)\equiv -\int_{0}^t\!\rmd t'\,\rme^{-(t-t')}\, \hat{d}_\mathrm{in}(t') ,
\end{equation}
which is simply the emission of a photon at time $t$ that could have been absorbed at anytime $t'$ from the start of the scattering event. The probability of absorption appears here in the form of the atomic exponential response function, which ensures that the photon is most likely to have been absorbed in a time window of order $\gamma^{-1}$ before the emission. As a side comment, the minus sign corresponds to the well-known $\pi$ phase shift from a dipole emission~\cite{Zumofen2008}.

It is important to note here that the linear form \eref{eq:sigma-Lin} is \emph{not} derived with the use of the weak-excitation limit, where one would typically assume that the atom mostly stays in the ground state and $\hat{\sigma}_z\to -1$ is set by hand. Instead, we use the argument that once $\hat{d}_\mathrm{in}$ has removed from the initial state a photon being scattered in the linear regime, the subsequent atomic operator $\hat{\sigma}_z$ in \eref{eq:sigma-} effectively acts on the empty state $\ket{\mathrm{vac}}\otimes\ket{g}$, yielding $-1$. In fact, one way to naturally visualize the linear regime is to imagine that the incoming photons are being sent on a set of $N$ colocated atoms, where each atom only sees one photon. This pictures that the atomic response to each photon is dictated by the single-photon regime, which is the essence of the linear regime, while no assumption is being made on the excitation of the atom.

Coming back to the actual response of the system~\eref{eq:sigma-}, it is therefore the presence of $\hat{\sigma}_z(t')$ that translates the nature of the atom as a saturable nonlinear scatterer and imposes, somehow, that the atom can only absorb or emit at most one photon at a given time. In the following, we will prove this statement and show how to use it to our advantage when describing a scattering event.

\section{Demonstrating the method in the 2-photon case}
In this section we will restrict ourselves to the case of a two-photon pulse incoming onto the atom initially in the ground state, that is $\ket{\psi_\mathrm{in}}=\ket{\psi_2}\otimes\ket{g}$. Here the goal is to introduce our novel approach in this preliminary situation before proceeding to the general case.

In a scattering event, one is typically interested in the evolution of wavepackets as a function of time. In particular, the L-S formalism gives access to the long-time limit of these wavepackets~\cite{Shen2007,Shen2015,Anders2015,Lee2015}. To illustrate our method, let us study the wavepacket associated with counter-propagating photons $f_1(\tau_1,\tau_2,t)\equiv\bra{\varnothing}\hat{a}^{\leftharpoonup}_{\tau_2}(t)\,\hat{a}^{\rightharpoonup}_{\tau_1}(t)\ket{\psi_\mathrm{in}}$, where $\ket{\varnothing}=\ket{\mathrm{vac}}\otimes\ket{g}$ and the initial state~\eref{eq:lightIn} implies $f_1(\tau_1,\tau_2,0)=\xi_1(\tau_1,\tau_2)$. Note that this does not restrict the class of input states $\ket{\psi_\mathrm{in}}$ but is merely a choice of what output we wish to look at, here being coincidence events. The study of the other wavepackets follows the same procedure.

\begin{figure}\centering
\subfloat{
\includegraphics[width=0.2\textwidth]{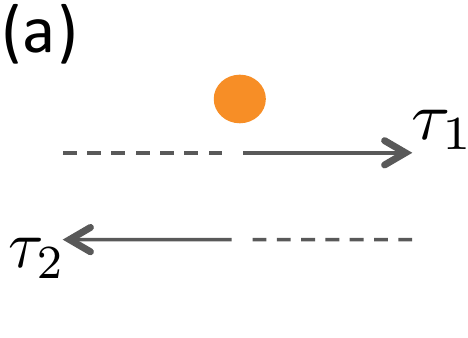}
}\quad
\subfloat{
\includegraphics[width=0.2\textwidth]{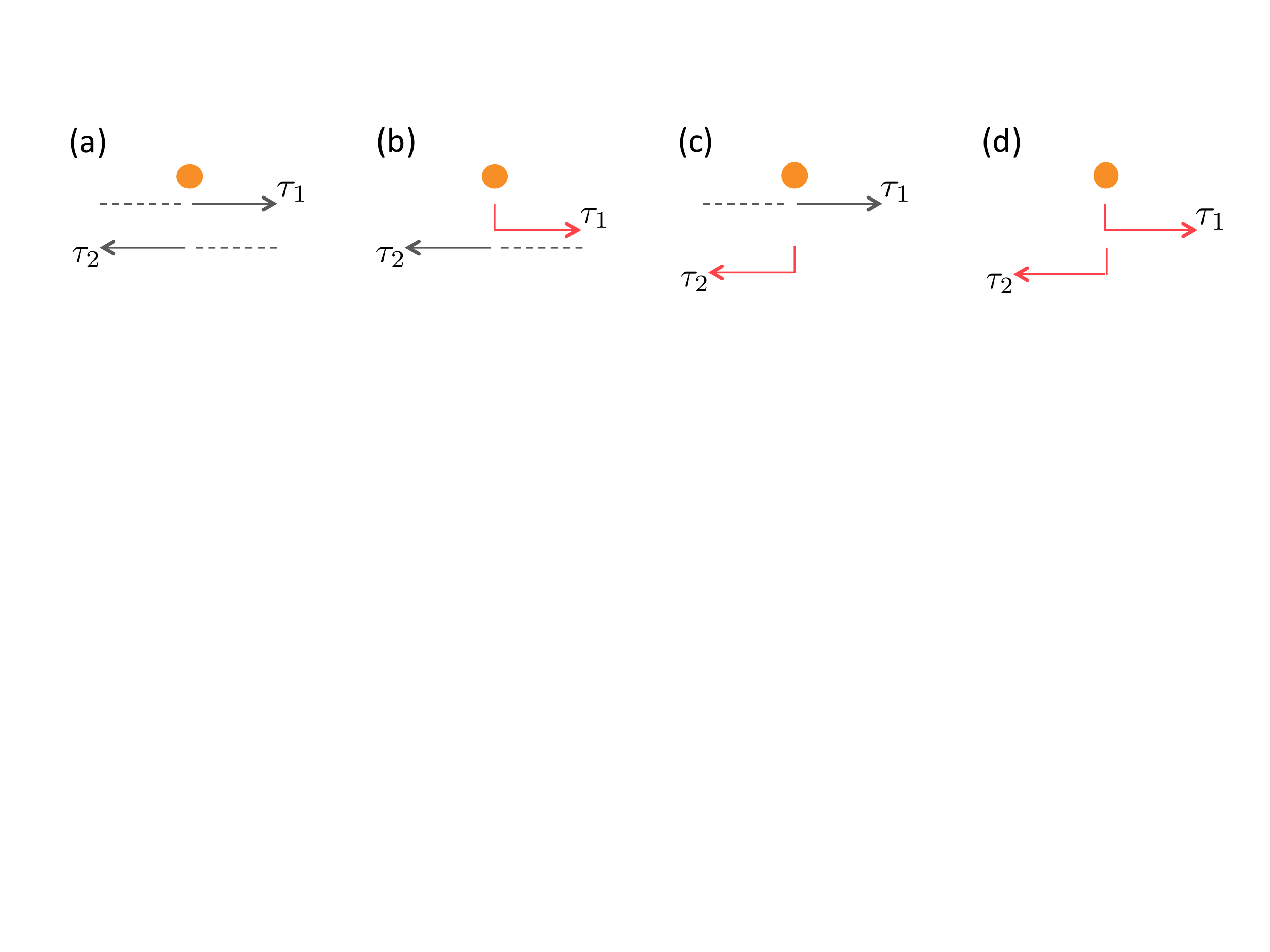}
}\quad
\subfloat{
\includegraphics[width=0.2\textwidth]{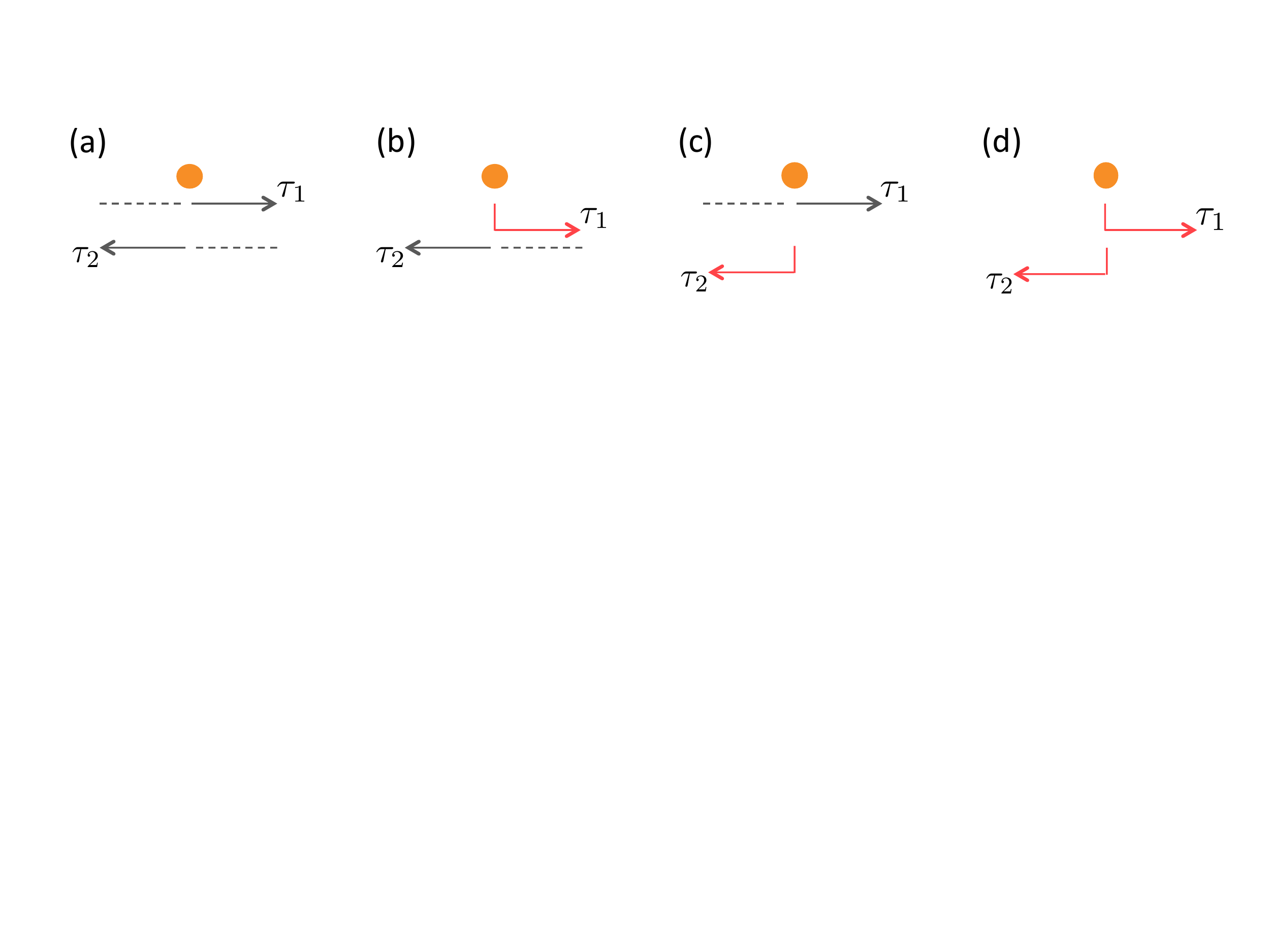}
}\quad
\subfloat{
\includegraphics[width=0.2\textwidth]{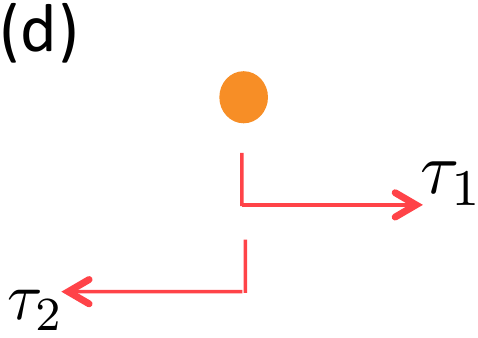}
}
\caption{\label{fig:diagramsPh}The different diagrams corresponding to the case of counter-propagating photons. (a) Both photons did not interact with the atom. (b) and (c) Only one of the photons has been absorbed and reemitted. (d) Both photons have been emitted by the atom.}
\end{figure}

\subsection{Step 1: drawing Feynman diagrams}
First, we start by decomposing the possibilities of having counter-propagating photons in terms of Feynman diagrams. Specifically, these diagrams correspond to the possible photonic \emph{paths} that shall be summed in order to obtain the amplitude of the process under study~\cite{Feynman1948,Roulet2016}. Note that, contrary to particle physics~\cite{Feynman1949}, our system does not call for a perturbative expansion as we can exhaustively list all the contributing diagrams. This is a direct consequence of the fact that each photon is only given a single chance to interact with the atom. Moreover, the strong directionality of the waveguide implies that the atom is effectively a two-input-two-output scatterer \cite{Roulet2016}.

The diagrams are shown in~\fref{fig:diagramsPh} and translate as
\begin{eqnarray}\label{eq:counterprop}\nonumber
	\hat{a}^{\leftharpoonup}_{\tau_2}(t)\,\hat{a}^{\rightharpoonup}_{\tau_1}(t)&=&\hat{a}^{\leftharpoonup}_{\tau_2}(0)\,\hat{a}^{\rightharpoonup}_{\tau_1}(0)
+\theta(t-\tau_1)\,\hat{\sigma}_-(\tau_1)\,\hat{a}^{\leftharpoonup}_{\tau_2}(0)
+\theta(t-\tau_2)\,\hat{\sigma}_-(\tau_2)\,\hat{a}^{\rightharpoonup}_{\tau_1}(0)\\
&+&\theta(t-\tau_1)\theta(t-\tau_2) \mathcal{T}\Big[\hat{\sigma}_-(\tau_2)\,\hat{\sigma}_-(\tau_1)\Big] ,
\end{eqnarray}
where the Heaviside functions ensure that a photon that has not yet reached the atom cannot originate from an atomic emission. Note that another straightforward way of obtaining this decomposition would be to substitute the solution of~\eref{eq:aDot}.

The first thing to notice at this stage is that all the terms in the first line of~\eref{eq:counterprop} are trivially linear and contribute to $f_1(\tau_1,\tau_2,t)$ as simple functions of the incoming wavepackets
\begin{eqnarray}\label{eq:triviallyLin}
	\bra{\varnothing}\hat{a}^{\leftharpoonup}_{\tau_2}(0)\,\hat{a}^{\rightharpoonup}_{\tau_1}(0)\ket{\psi_\mathrm{in}}=\xi_1(\tau_1,\tau_2) ,\\\nonumber
	\bra{\varnothing}\hat{\sigma}^\mathrm{lin}_-(\tau_1)\,\hat{a}^{\leftharpoonup}_{\tau_2}(0)\ket{\psi_\mathrm{in}}=-\int_{0}^{\tau_1}\!\rmd t'\,\rme^{-(\tau_1-t')} \Big[\sqrt{2}\,\xi_0(t',\tau_2)+\xi_1(t',\tau_2)\Big],\\\nonumber
	\bra{\varnothing}\hat{\sigma}^\mathrm{lin}_-(\tau_2)\,\hat{a}^{\rightharpoonup}_{\tau_1}(0)\ket{\psi_\mathrm{in}}=-\int_{0}^{\tau_2}\!\rmd t'\,\rme^{-(\tau_2-t')} \Big[\sqrt{2}\,\xi_2(\tau_1,t')+\xi_1(\tau_1,t')\Big] .
\end{eqnarray}
Therefore in the process of solving the scattering of two photons, it is the last term of~\eref{eq:counterprop}, which involves two successive atomic emissions, that calls for a more thorough analysis.

\subsection{Step 2: treating the atom as a non-saturable linear beamsplitter with a twist}
Let us assume here that $\tau_2$ is greater than $\tau_1$ without loss of generality. If the atom was a linear scatterer~\eref{eq:sigma-Lin}, the last term in~\eref{eq:counterprop} would read
\begin{eqnarray}\label{eq:ifLinearBS}
	\bra{\varnothing}\hat{\sigma}^\mathrm{lin}_-(\tau_2)\,\hat{\sigma}^\mathrm{lin}_-(\tau_1)&&\ket{\psi_\mathrm{in}}=\int_{0}^{\tau_2}\!\rmd t''\,\int_{0}^{\tau_1}\!\rmd t'\,\rme^{-(\tau_2-t'')}\rme^{-(\tau_1-t')}\cdot\\\nonumber
	&&\Big[\sqrt{2}\,\xi_0(t',t'')+\xi_1(t',t'')+\xi_1(t'',t')+\sqrt{2}\,\xi_2(t',t'')\Big] .
\end{eqnarray}
Now our main result is that the actual atomic response is in fact of the form 
\begin{equation}\label{eq:main2ph}
	\bra{\varnothing}\hat{\sigma}_-(\tau_2)\,\hat{\sigma}_-(\tau_1)\ket{\psi_\mathrm{in}}=\bra{\varnothing}\hat{\sigma}^\mathrm{lin}_-(\tau_2,\tau_1)\,\hat{\sigma}^\mathrm{lin}_-(\tau_1)\ket{\psi_\mathrm{in}} ,
\end{equation}
where
\begin{equation}
	\hat{\sigma}^{\mathrm{lin}}_-(\tau_2,\tau_1)\equiv -\int_{\tau_1}^{\tau_2}\!\rmd t'\,\rme^{-(\tau_2-t')}\, \hat{d}_\mathrm{in}(t') ,
\end{equation}
takes into account the fact that the photon emitted at $\tau_2$ has necessarily been absorbed after the emission of the first photon at $\tau_1$. Therefore, in practice, one simply needs to shift the start of the first integral in the linear expression~\eref{eq:ifLinearBS} in order to fully grasp the impact of the atomic nonlinearity onto the scattered light \emph{without doing any computation}. This simple substitution concludes the derivation of $f_1(\tau_1,\tau_2,t)$ which is now expressed solely in terms of simple integrals over the incoming wavepackets (in Appendix B we show that the results of~\cite{Shen2007A} derived via L-S are recovered by our method).

\subsection{Proof of the main result in the two-photon case}
We will now present a detailed proof of the main result~\eref{eq:main2ph} for the input state $\ket{\psi_\mathrm{in}}=\ket{\psi_2}\otimes\ket{g}$. We start by expressing the nonlinear term using~\eref{eq:sigma-}
\begin{equation}
	\bra{\varnothing}\hat{\sigma}_-(\tau_2)\,\hat{\sigma}_-(\tau_1)\ket{\psi_\mathrm{in}}= \int_{0}^{\tau_1}\!\rmd t'\,\rme^{-(\tau_1-t')} \bra{\varnothing}\hat{\sigma}_-(\tau_2)\,\hat{\sigma}_z(t')\,\hat{d}_\mathrm{in}(t')\ket{\psi_\mathrm{in}},
\end{equation}
where we recall that $\tau_2$ is assumed to be greater than $\tau_1$.

As pointed out in \sref{sec:model}, it is the presence of $\hat{\sigma}_z(t')$ that dictates the nonlinear response of the atom. However at this level it is not clear yet how to interpret its role. In order to proceed further, we will use the solution of~\eref{eq:sigmazDot}
\begin{equation}
	\hat{\sigma}_z(t')=\rme^{-2t'}\,\Big[\hat{\mathds{1}}+\hat{\sigma}_z(0)\Big]- \mathds{1}-2\int_{0}^{t'}\!\rmd t''\, \rme^{-2(t'-t'')}\, \Big[\hat{\sigma}_+(t'')\, \hat{d}_\mathrm{in}(t'')+\mathrm{H.c}\Big] .
\end{equation}
This gives us four contributions as follows
\begin{itemize}
	\item $\int_{0}^{\tau_1}\!\rmd t'\,\rme^{-(\tau_1-t')}\rme^{-2t'} \bra{\varnothing}\hat{\sigma}_-(\tau_2)\,\Big[\hat{\mathds{1}}+\hat{\sigma}_z(0)\Big]\,\hat{d}_\mathrm{in}(t')\ket{\psi_\mathrm{in}}=0$ where we used the absence of initial excitation in the atom;\newline
	\item $-\int_{0}^{\tau_1}\!\rmd t'\,\rme^{-(\tau_1-t')} \bra{\varnothing}\hat{\sigma}_-(\tau_2)\,\hat{d}_\mathrm{in}(t')\ket{\psi_\mathrm{in}}=\bra{\varnothing}\hat{\sigma}^\mathrm{lin}_-(\tau_2)\,\hat{\sigma}^\mathrm{lin}_-(\tau_1)\ket{\psi_\mathrm{in}}$ which corresponds to a linear scatterer~\eref{eq:ifLinearBS}. The remaining terms will thus contain the nonlinear correction;\newline
	\item \quad $-2\int_{0}^{\tau_1}\!\rmd t'\,\int_{0}^{t'}\!\rmd t''\,\rme^{-(\tau_1-t')}\rme^{-2(t'-t'')} \bra{\varnothing}\hat{\sigma}_-(\tau_2)\,\hat{\sigma}_+(t'')\, \hat{d}_\mathrm{in}(t'')\,\hat{d}_\mathrm{in}(t')\ket{\psi_\mathrm{in}}$\newline $=-2\int_{0}^{\tau_1}\!\rmd t'\,\int_{0}^{t'}\!\rmd t''\,\rme^{-(\tau_1+\tau_2)}\rme^{-(t'-3t'')} \bra{\varnothing} \hat{d}_\mathrm{in}(t'')\,\hat{d}_\mathrm{in}(t')\ket{\psi_\mathrm{in}}$ where the simplification arises from the presence of only two photons in the input state. The atomic operators are thus effectively acting on the vacuum state;\newline
	\item \quad $-2\int_{0}^{\tau_1}\!\rmd t'\,\int_{0}^{t'}\!\rmd t''\,\rme^{-(\tau_1-t')}\rme^{-2(t'-t'')} \bra{\varnothing}\hat{\sigma}_-(\tau_2)\,\hat{d}^\dagger_\mathrm{in}(t'')\,\hat{\sigma}_-(t'')\,\hat{d}_\mathrm{in}(t')\ket{\psi_\mathrm{in}}$\newline $=-4\int_{0}^{\tau_1}\!\rmd t'\,\int_{0}^{t'}\!\rmd t''\,\int_{0}^{t''}\!\rmd t'''\,\rme^{-(\tau_1+\tau_2)}\rme^{-(t'-2t''-t''')} \bra{\varnothing} \hat{d}_\mathrm{in}(t''')\,\hat{d}_\mathrm{in}(t')\ket{\psi_\mathrm{in}}$ where we have again used the presence of only two photons in the input state and substituted $\hat{\sigma}_-(t'')$ using~\eref{eq:sigma-Dot}, yielding a third integral.\newline
\end{itemize}
The final step consists in changing the order of integration between $t''$ and $t'''$ in the last contribution $\int_{0}^{t'}\!\rmd t''\,\int_{0}^{t''}\!\rmd t'''\to\int_{0}^{t'}\!\rmd t'''\,\int_{t'''}^{t'}\!\rmd t''$ so that the integral over $t''$, which does not involve any operator, can be evaluated. One is then left with a term that cancels the third contribution plus an additional term, the nonlinear correction, which reads
\begin{eqnarray}\label{eq:nonlinCorrection}
	-2&&\int_{0}^{\tau_1}\!\rmd t'\,\int_{0}^{t'}\!\rmd t''\,\rme^{-(\tau_1+\tau_2)}\rme^{-(t'+t'')} \bra{\varnothing} \hat{d}_\mathrm{in}(t'')\,\hat{d}_\mathrm{in}(t')\ket{\psi_\mathrm{in}} \\\nonumber
	&&=-\int_{0}^{\tau_1}\!\rmd t'\,\int_{0}^{\tau_1}\!\rmd t''\,\rme^{-(\tau_1+\tau_2)}\rme^{-(t'+t'')} \bra{\varnothing} \hat{d}_\mathrm{in}(t'')\,\hat{d}_\mathrm{in}(t')\ket{\psi_\mathrm{in}} .
\end{eqnarray}

The role of this contribution is now transparent and yields the main result~\eref{eq:main2ph} when combined with the linear term~\eref{eq:ifLinearBS}.

\subsection{Atomic excitation during the scattering event}
In fact, our method is not limited to describing the effect of the atom onto the light but also proves useful for instance to track the atomic excitation during the scattering event. Indeed, the probability of excitation is given by
\begin{eqnarray}
	P_e(t)&=&\bra{\psi_\mathrm{in}}\hat{\sigma}_+(t)\,\hat{\sigma}_-(t)\ket{\psi_\mathrm{in}}\\\nonumber
	&=&\int_{0}^\infty\!\rmd \tau\,|\bra{\varnothing}\hat{\sigma}_-(t)\,\hat{a}^{\rightharpoonup}_{\tau}(t)\ket{\psi_\mathrm{in}}|^2+|\bra{\varnothing}\hat{\sigma}_-(t)\,\hat{a}^{\leftharpoonup}_{\tau}(t)\ket{\psi_\mathrm{in}}|^2 ,
\end{eqnarray}
which corresponds to an excitation being present in the atom at time $t$ while the other photon is propagating in any direction.

\begin{figure}\centering
\subfloat{
\includegraphics[width=0.2\textwidth]{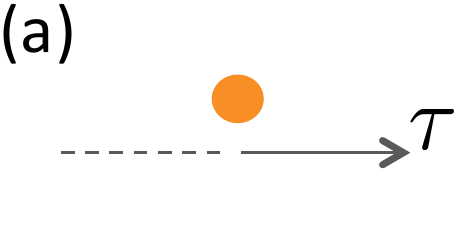}
}\quad
\subfloat{
\includegraphics[width=0.2\textwidth]{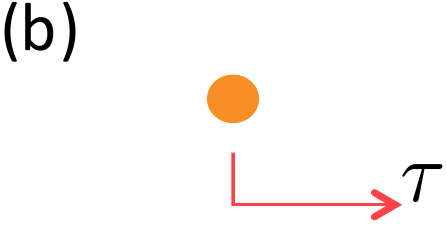}
}
\caption{\label{fig:diagramsExc}The different diagrams corresponding to a photon propagating to the right. (a) This photon did not interact with the atom. (b) It has been absorbed and reemitted.}
\end{figure}

Following step 1, we first express the photon field in terms of Feynman diagrams as illustrated in~\fref{fig:diagramsExc}
\begin{eqnarray}\nonumber
	\hat{\sigma}_-(t)\,\hat{a}^{\rightharpoonup}_{\tau}(t)&=&\hat{\sigma}_-(t)\,\hat{a}^{\rightharpoonup}_{\tau}(0)+\theta(t-\tau)\,\hat{\sigma}_-(t)\,\hat{\sigma}_-(\tau) ,
\end{eqnarray}
and similarly for $\hat{\sigma}_-(t)\,\hat{a}^{\leftharpoonup}_{\tau}(t)$.
Again the first term's contribution is trivially linear and is readily given in terms of the initial wavepackets as shown in~\eref{eq:triviallyLin}. On the other hand, we recognize the second term and applying our main result~\eref{eq:main2ph} ends the derivation.

\section{Main result in the case of $N$ initial excitations}
Having established the working principle of our method in the two-photon case, we will now proceed to the most general situation where the number of incoming photons is arbitrary and the atom is not necessarily in the ground state. Specifically, the input state consists of $N$ initial excitations distributed between the field and the atom $\ket{\psi_\mathrm{in}}= c_g\ket{\psi_N}\otimes\ket{g}+c_e\ket{\psi_{N-1}}\otimes\ket{e}$.

\subsection{An intuitive generalization}
When decomposing in terms of Feynman diagrams, the possibility of having multiple photons originating from atomic emissions gives rise to higher-order contributions. These are responsible for the complexity of previous approaches which treat the atom as a generic scatterer and require to solve the scattering problem for each of these diagrams from scratch, recalculating the system's response. Here lies the advantage of our method, which is based on the physics of the atomic response and allows to bypass calculations by capturing the system's nonlinear response in a transparent substitution. Indeed, our main result extends naturally to the case of $N$ excitations
\begin{eqnarray}\label{eq:nExc}
	\bra{\varnothing}&\hat{\sigma}_-&(\tau_N)\cdots\hat{\sigma}_-(\tau_1)\ket{\psi_\mathrm{in}}\\\nonumber
	&=& \bra{\varnothing}\hat{\sigma}^\mathrm{lin}_-(\tau_N,\tau_{N-1})\cdots\hat{\sigma}^\mathrm{lin}_-(\tau_2,\tau_1)\,\Big[\rme^{-\tau_1}\hat{\sigma}_-(0)\,+\,\hat{\sigma}^\mathrm{lin}_-(\tau_1)\Big]\ket{\psi_\mathrm{in}} ,
\end{eqnarray}
where $\tau_N>\ldots>\tau_1$ are time-ordered. This is a remarkably simple and intuitive \emph{operational translation} of the atomic nonlinearity, which allows one to derive any quantity of interest as straightforwardly as if the atom was a linear beamsplitter. Moreover, the fact that the scatterer could have some initial excitation before the scattering event is nicely taken into account by the term $\rme^{-\tau_1}\hat{\sigma}_-(0)$.

\subsection{Proof of the main result in the $N$-excitation case}
In this section we will prove the main result in the $N$-excitation manifold~\eref{eq:nExc} by induction based on the $2$-excitation result. For the latter, in addition to~\eref{eq:main2ph} we also need to consider the case where the atom is initially excited, which yields (we will assume $\tau_{i+1}>\tau_i$ in the following)
\begin{eqnarray}
	\bra{\varnothing}&&\hat{\sigma}_-(\tau_2)\,\hat{\sigma}_-(\tau_1)\Big(\ket{\psi_1}\otimes\ket{e}\Big)\\\nonumber
	&&= \rme^{-\tau_1}\bra{\varnothing}\hat{\sigma}_-(\tau_2)\Big(\ket{\psi_1}\otimes\ket{g}\Big)+\int_{0}^{\tau_1}\!\rmd t'\,\rme^{-(\tau_1-t')} \bra{\varnothing}\hat{\sigma}_-(\tau_2)\,\hat{\sigma}_z(t')\,\hat{d}_\mathrm{in}(t')\Big(\ket{\psi_1}\otimes\ket{e}\Big)\\\nonumber
	&&= \rme^{-\tau_1}\Big[\bra{\varnothing}\hat{\sigma}^\mathrm{lin}_-(\tau_2)\Big(\ket{\psi_1}\otimes\ket{g}\Big)+\int_{0}^{\tau_1}\!\rmd t'\,\rme^{-(\tau_2-t')} \bra{\varnothing}\hat{d}_\mathrm{in}(t')\Big(\ket{\psi_1}\otimes\ket{g}\Big)\Big]\\\nonumber
	&&= \rme^{-\tau_1} \bra{\varnothing}\hat{\sigma}^\mathrm{lin}_-(\tau_2,\tau_1)\Big(\ket{\psi_1}\otimes\ket{g}\Big) .
\end{eqnarray}

The base for our induction proof thus reads
\begin{equation}\label{eq:2ExcMan}
	\bra{\varnothing}\hat{\sigma}_-(\tau_2)\,\hat{\sigma}_-(\tau_1)\ket{2\mathrm{\ exc.}}=\bra{\varnothing}\hat{\sigma}^\mathrm{lin}_-(\tau_2,\tau_1)\,\Big[\rme^{-\tau_1}\hat{\sigma}_-(0)\,+\,\hat{\sigma}^\mathrm{lin}_-(\tau_1)\Big]\ket{2\mathrm{\ exc.}} .
\end{equation}
where $\ket{2\mathrm{\ exc.}}$ stands for any initial state consisting of two excitations distributed between the light field and the atom.

Now let us assume that our result holds in the $N$-excitation manifold
\begin{eqnarray}
	\bra{\varnothing}&\hat{\sigma}_-&(\tau_N)\cdots\hat{\sigma}_-(\tau_1)\ket{N\mathrm{\ exc.}}\\\nonumber
	&=& \bra{\varnothing}\hat{\sigma}^\mathrm{lin}_-(\tau_N,\tau_{N-1})\cdots\hat{\sigma}^\mathrm{lin}_-(\tau_2,\tau_1)\,\Big[\rme^{-\tau_1}\hat{\sigma}_-(0)\,+\,\hat{\sigma}^\mathrm{lin}_-(\tau_1)\Big]\ket{N\mathrm{\ exc.}} ,
\end{eqnarray}
and prove that this implies that our result also holds in the $N+1$-excitation manifold. This reads
\begin{eqnarray}
	\bra{\varnothing}&\hat{\sigma}_-&(\tau_{N+1})\cdots\hat{\sigma}_-(\tau_1)\ket{N+1\mathrm{\ exc.}}\\\nonumber
	&=& \bra{\varnothing}\hat{\sigma}^\mathrm{lin}_-(\tau_{N+1},\tau_N)\cdots\hat{\sigma}^\mathrm{lin}_-(\tau_3,\tau_2)\,\Big[\rme^{-\tau_2}\hat{\sigma}_-(0)\,+\,\hat{\sigma}^\mathrm{lin}_-(\tau_2)\Big]\,\hat{\sigma}_-(\tau_1)\ket{N+1\mathrm{\ exc.}}\\\nonumber	 	&=&\bra{\varnothing}\hat{\sigma}_-(\tau_2)\,\hat{\sigma}_-(\tau_1)\,\underbrace{\hat{\sigma}^\mathrm{lin}_-(\tau_{N+1},\tau_N)\cdots\hat{\sigma}^\mathrm{lin}_-(\tau_3,\tau_2)}_{\mathrm{removes\ }N-1\mathrm{\ photons\ from\ initial\ state}}\ket{N+1\mathrm{\ exc.}} ,
\end{eqnarray}
where we used the commutation relation $\Big[\hat{\sigma}^\mathrm{lin}_-(\tau_{i+1},\tau_i)\,,\,\hat{\sigma}_-(\tau_1)\Big]=0$ for any $\tau_i>\tau_1$. Now applying the previous result on the 2-excitation manifold~\eref{eq:2ExcMan} and rearranging the terms concludes the proof and one finds
\begin{equation}
	\bra{\varnothing}\hat{\sigma}^\mathrm{lin}_-(\tau_{N+1},\tau_N)\cdots\hat{\sigma}^\mathrm{lin}_-(\tau_2,\tau_1)\,\Big[\rme^{-\tau_1}\hat{\sigma}_-(0)\,+\,\hat{\sigma}^\mathrm{lin}_-(\tau_1)\Big]\ket{N+1\mathrm{\ exc.}} .
\end{equation}

\section{Solving the reflection of $N$ photons}
In order to illustrate the power of our method, we will go beyond the usual few-photon examples addressed in the literature and tackle the reflection of $N$ photons on the atom initially in the ground state $\ket{\psi_\mathrm{in}}=\ket{\psi_{N}}\otimes\ket{g}$. For the sake of concreteness, we will assume that these photons are all incoming from the left, such that only the wavepacket $\xi_{N}(\tau_1,\ldots,\tau_{N})$ contributes to the input state~\eref{eq:lightIn}. The most complete description of the reflection event is given by the wavepacket corresponding to all the photons traveling to the left $f_0(\tau_1,\ldots,\tau_{N},t)\equiv \bra{\varnothing}\hat{a}^{\leftharpoonup}_{\tau_{N}}(t)\cdots\hat{a}^{\leftharpoonup}_{\tau_1}(t)\ket{\psi_\mathrm{in}}/\sqrt{N!}$ with $f_0(\tau_1,\ldots,\tau_{N},0)=0$ at initial time.

When decomposing in terms of Feynman diagrams, it is clear that all the $N$ photons have to be absorbed by the atom in order to reverse their direction of propagation. Therefore, the first step reads
\begin{equation}\label{eq:Nref}
	\bra{\varnothing}\hat{a}^{\leftharpoonup}_{\tau_{N}}(t)\cdots\hat{a}^{\leftharpoonup}_{\tau_1}(t)\ket{\psi_\mathrm{in}}=\theta(t-\tau_1)\cdots\theta(t-\tau_{N}) \bra{\varnothing}\mathcal{T}\Big[\hat{\sigma}_-(\tau_{N})\cdots\hat{\sigma}_-(\tau_1)\Big]\ket{\psi_\mathrm{in}} ,
\end{equation}
where we used our knowledge of the initial state to dramatically reduce the number of contributing Feynman diagrams to 1. From now on, let us assume $\tau_N>\ldots>\tau_1$ are time-ordered without loss of generality. We can then apply our main result, which gives
\begin{eqnarray}\label{eq:mainResultRef}
	\!\!\!\!\!\!\!\!\!\bra{\varnothing}&\hat{\sigma}_-&(\tau_{N})\cdots\hat{\sigma}_-(\tau_1)\ket{\psi_\mathrm{in}}=\bra{\varnothing}\hat{\sigma}^\mathrm{lin}_-(\tau_{N},\tau_{N-1})\cdots\hat{\sigma}^\mathrm{lin}_-(\tau_2,\tau_1)\,\hat{\sigma}^\mathrm{lin}_-(\tau_1)\ket{\psi_\mathrm{in}}\\\nonumber
	&=&(-1)^N\sqrt{N!}\,\int_{\tau_{N-1}}^{\tau_{N}}\!\rmd t_{N}\,\cdots\int_{\tau_1}^{\tau_{2}}\!\rmd t_2\,\int_{0}^{\tau_{1}}\!\rmd t_1\,\rme^{-(\tau_{N}-t_{N})}\cdots\rme^{-(\tau_1-t_1)}\,\xi_{N}(t_1,\ldots,t_{N}) ,
\end{eqnarray}
and ends the derivation of $f_0(\tau_1,\ldots,\tau_N,t)$ without a single calculation being done. Intuitively, we are asking all the photons to be absorbed subsequently given their initial distribution $\xi_{N}(t_1,\ldots,t_{N})$.

As a practical example, we will consider the specific incoming light that would be generated from the proposal~\cite{Gonzalez2015}. There, the authors show how to use a large number of three-level atoms in a superradiant configuration as a source for deterministically generating $N$-photon states in the 1D waveguide. In the simplest case, the photons would all be emitted at $\omega_A$ with the same exponentially decaying profile $\xi_{N}(\tau_1,\ldots,\tau_{N})=\prod^{N}_{i=1} \xi(\tau_i)$ where
\begin{equation}
	\xi(\tau_i)= \sqrt{\Gamma}\,\rme^{-\tau_i\,\Gamma/2} .
\end{equation}
Note that the frequency bandwidth of this mode $\Gamma$ is essentially given by the superradiant decay rate of the source and is therefore adjustable by adding or removing atoms from the source. Since these photons are all emitted into the same mode, it is straightforward to show using \eref{eq:mainResultRef} that $f_0(\tau_1,\ldots,\tau_N,t)$ is given by a product of integrals of the form
\begin{equation}
	h(\tau_{i},\tau_{i-1})\equiv\int_{\tau_{i-1}}^{\tau_{i}}\!\rmd t_i\,\rme^{-(\tau_{i}-t_{i})}\,\xi(t_i)=\sqrt{\Gamma}\,\frac{\rme^{- \tau_i\,\Gamma/2}-\rme^{-\tau_i+\tau_{i-1}\,(1-\Gamma/2)}}{1-\Gamma/2} .
\end{equation}

From the knowledge of $f_0(\tau_1,\ldots,\tau_N,t)$, we also have access to the probability of finding all the photons reflected after the scattering event $R_N$. Indeed, the latter is naturally obtained by integrating the former in the long-time limit as follows
\begin{eqnarray}\label{eq:RN}
	R_N&=&\lim\limits_{t\to\infty}\,\int_{0}^{\infty}\!\rmd \tau_{N}\,\cdots\int_{0}^{\infty}\!\rmd \tau_1\,|f_0(\tau_1,\ldots,\tau_N,t)|^2\\\nonumber
	&=& N!\,\lim\limits_{t\to\infty}\,\int_{0}^{\infty}\!\rmd \tau_{1}\int_{\tau_{1}}^{\infty}\!\rmd \tau_{2}\,\cdots\int_{\tau_{N-1}}^{\infty}\!\rmd \tau_N\,|f_0(\tau_1,\ldots,\tau_N,t)|^2\\\nonumber
	&=& N!\,\int_{0}^{\infty}\!\rmd \tau_{1}\,|h(\tau_1,0)|^2\int_{\tau_{1}}^{\infty}\!\rmd \tau_{2}\,|h(\tau_2,\tau_1)|^2\cdots\int_{\tau_{N-1}}^{\infty}\!\rmd \tau_N\,|h(\tau_N,\tau_{N-1})|^2 ,
\end{eqnarray}
where we have used the bosonic exchange symmetry to rearrange the integral in a time-ordered manner. Also note that the dynamical time $t$ only appears via the Heaviside functions in \eref{eq:Nref}, which are equal to unity in the long-time limit.

\begin{figure}
\includegraphics[width=0.6\textwidth]{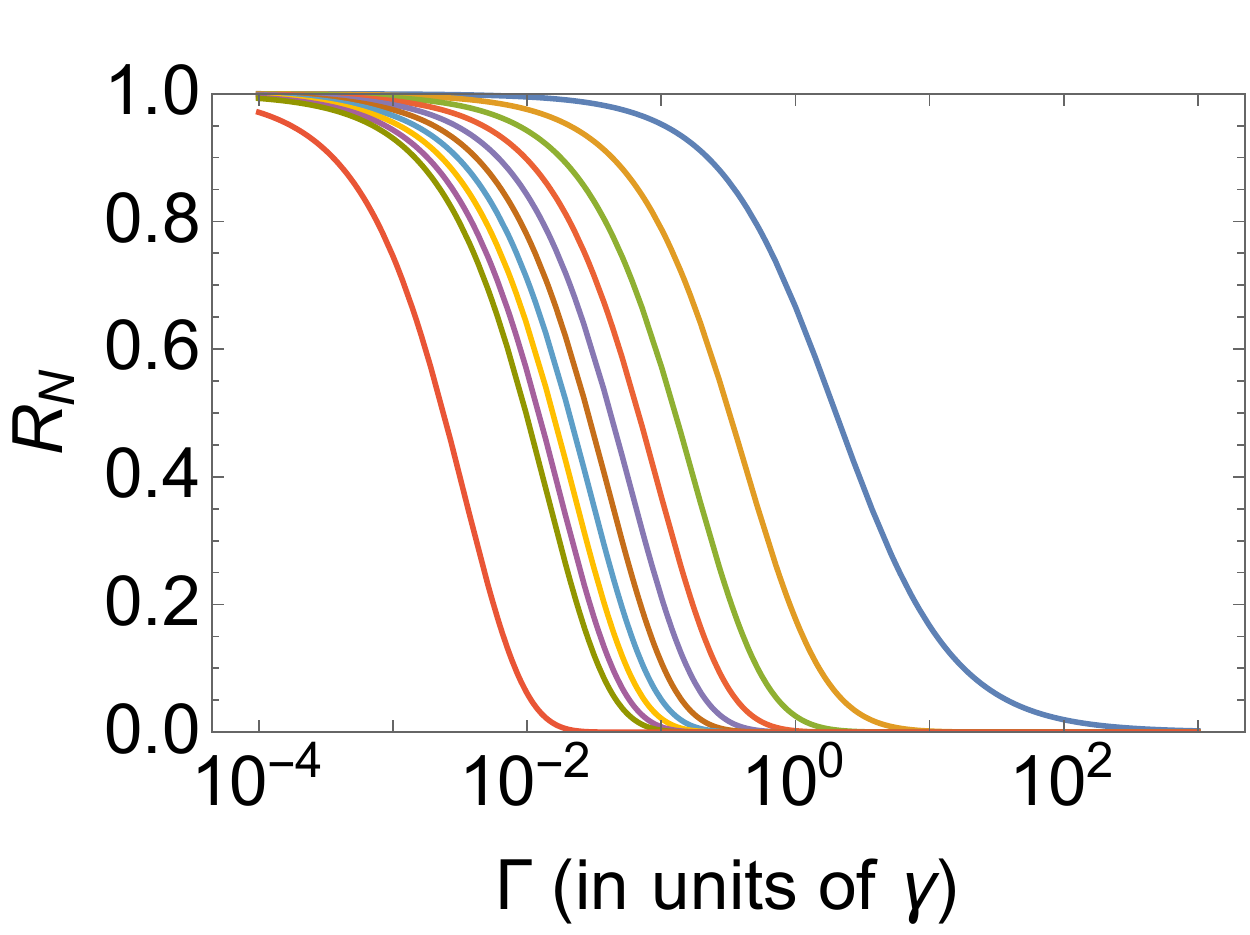}\centering
\caption{\label{fig:Nref}Probability of finding the $N$ photons reflected as a function of the frequency bandwidth $\Gamma$ normalized to the atomic linewidth $\gamma$. The curves from right to left correspond to a photon number $N$ of $\{ 1,2,3,4,5,6,7,8,9,10,20 \}$.}
\end{figure}

Finally, the expression for $R_N$ can be given a compact analytical form by noticing the following property for any $m\in\mathbb{N}$
\begin{equation}
	\int_{\tau_{i-1}}^{\infty}\!\rmd \tau_{i}\,\rme^{-m\,\tau_i\,\Gamma}\,|h(\tau_i,\tau_{i-1})|^2= \frac{4\,\rme^{-(1+m)\,\tau_{i-1}\,\Gamma}}{(1+m)(2+m\Gamma)(2+\Gamma+2m\Gamma)} ,
\end{equation}
which implies that the probability is given by
\begin{equation}
	R_N=N!\,\prod_{m=0}^{N-1}\frac{4}{(1+m)(2+m\Gamma)(2+\Gamma+2m\Gamma)} .
\end{equation}
Therefore, using our method, we have obtained the probability of finding the $N$ photons reflected to the left as a function of their frequency bandwidth $\Gamma$. The result is given in~\fref{fig:Nref} for various $N$. Its qualitative features can be understood. In the limit of large bandwidth, more and more frequency components of the pulse are off-resonant and simply pass by the atom without interacting. This explains $R_N\to 0$ for $\Gamma\to\infty$. In the limit of very narrow bandwidth, the $N$-photon pulse is very long in time, thus the atom effectively responds to each photon separately; and for a single-photon at resonance, full reflection is expected~\cite{Shen2005}, whence $R_N\to 1$ for $\Gamma\to 0$. Between these two cases, the higher the number of photons, the faster $R_N$ drops. This is the manifestation of the fact that the atom can absorb only one photon at a time: once one is absorbed, the others pass through.

\section{Conclusion}
We have presented a novel approach for solving the scattering of $N$ photons on a two-level atom. At the heart of the method lies the operational translation of the atomic nonlinear response, which allows to derive any quantity of interest in terms of the initial wavepackets without effort. Importantly, the method goes beyond the long-time limit, giving access to quantities like the atomic excitation during the scattering event. This is especially relevant in the context of an experiment, where one would ideally want to fully characterize the dynamical evolution of the light field and the atom. We have also applied our method to a proposed protocol for reliably generating $N$-photon pulses in the laboratory. An interesting perspective would be to consider the case of coherent input light, as well as to extend the concept to a three-level atom, understanding how this new level structure would translate at the operational level.
\ack
We thank Dimitris~G.~Angelakis, Ewan~Munro and Stefan Nimmrichter for insightful discussions. This research is supported by the National Research Foundation (partly through its Competitive Research Programme, Award No. NRF-CRP12-2013-03) and the Ministry of Education, Singapore.

\appendix
\section{Detailed derivation of the Heisenberg equations}
\setcounter{section}{1}

We here provide additional details for the reader interested in deriving the set of equations (3). As mentioned in the main text, we will be working in the interaction picture. The Hamiltonian of the system therefore reads~\cite{Domokos2002}
\begin{equation}
	\hat{H}_{I}=-i\hbar\int_{0}^\infty\!\rmd \omega\, g_\omega \Big[\ket{e}\bra{g} \left(\hat{a}^\leftharpoonup_\omega +\hat{a}^\rightharpoonup_\omega \right)\rme^{-i(\omega-\omega_A)t}-\mathrm{H.c.} \Big] ,
\end{equation}
which is equivalent to the dipole Hamiltonian~\eref{eq:Hamiltonian} up to a time-dependent phase.

Using the standard procedure, we can then derive the Heisenberg equations as
\begin{eqnarray}
	\frac{\mathrm{d}}{\mathrm{d}t}\hat{\sigma}_z &=&\frac{i}{\hbar}\Big[\hat{H}_{I}\,,\,\hat{\sigma}_z\Big]= -2\int_{0}^\infty\!\rmd \omega\, g_\omega \Big[\hat{\sigma}_+(\hat{a}^\leftharpoonup_\omega+\hat{a}^\rightharpoonup_\omega)\rme^{-i(\omega-\omega_A)t}+\mathrm{H.c.} \Big] ,\label{A:Heiz1}\\
	\frac{\mathrm{d}}{\mathrm{d}t}\hat{\sigma}_- &=&\frac{i}{\hbar}\Big[\hat{H}_{I}\,,\,\hat{\sigma}_-\Big]= \hat{\sigma}_z\int_{0}^\infty\!\rmd \omega\, g_\omega (\hat{a}^\leftharpoonup_\omega+\hat{a}^\rightharpoonup_\omega)\rme^{-i(\omega-\omega_A)t} ,\label{A:Hei-1}\\
	\frac{\mathrm{d}}{\mathrm{d}t}\hat{a}^\rightharpoonup_\omega&=&\frac{i}{\hbar}\Big[\hat{H}_{I}\,,\,\hat{a}^\rightharpoonup_\omega\Big]=g_\omega\rme^{i(\omega-\omega_A)t}\, \hat{\sigma}_- ,\label{A:Heiar}\\
	\frac{\mathrm{d}}{\mathrm{d}t}\hat{a}^\leftharpoonup_\omega &=&\frac{i}{\hbar}\Big[\hat{H}_{I}\,,\,\hat{a}^\leftharpoonup_\omega\Big]=g_\omega\rme^{i(\omega-\omega_A)t}\, \hat{\sigma}_- .\label{A:Heial}
\end{eqnarray}
Equation \eref{eq:aDot} is then readily obtained by taking the Fourier transform of \eref{A:Heiar} and \eref{A:Heial} and neglecting the variation of the coupling constant around the transition frequency $g_\omega\approx g_{\omega_A}$. This latter approximation will be used in the rest of the calculations and is the standard Weisskopf-Wigner approximation, under which we can safely extend the frequency integration to $-\infty$~(see Chapter 6 of \cite{Scully1997}).

The next step consists in eliminating the time-dependent field operators from the atom equations \eref{A:Heiz1} and \eref{A:Hei-1}. To this end, we formally integrate the field operators as
\begin{equation}\label{A:formal}
	\hat{a}^\rightharpoonup_\omega(t)=\hat{a}^\rightharpoonup_\omega(0)+g_\omega\int_{0}^{t}\!\rmd t'\,\rme^{i(\omega-\omega_A)t'}\, \hat{\sigma}_-(t') ,
\end{equation}
and similarly for $\hat{a}^\leftharpoonup_\omega(t)$. Substituting this form into \eref{A:Heiz1} and \eref{A:Hei-1} gives rise to two types of contributions:
\begin{itemize}
	\item the first contribution comes from the field operators at dynamical time $t=0$ and yields the terms proportional to $\hat{d}_\mathrm{in}$ and $\hat{d}_\mathrm{in}^\dagger$ in \eref{eq:sigmazDot} and \eref{eq:sigma-Dot}. These are field operators whose action on the initial state is well-defined;
	\item substituting the second term of \eref{A:formal} yields double integrals. Performing first the integral over frequency gives a delta function in time $\int_{-\infty}^\infty\!\rmd\omega\,\rme^{i(\omega-\omega_A)(t'-t)}=2\pi\delta (t'-t)$ which can be evaluated to obtain \eref{eq:sigmazDot} and \eref{eq:sigma-Dot}.
\end{itemize}

Note that an even more straightforward derivation is possible by noticing that the field operators already appear as Fourier transforms in \eref{A:Heiz1} and \eref{A:Hei-1} under the Weisskopf-Wigner approximation.

\section{Comparison with results obtained via L-S formalism}
In this section we will consider the two-photon scattering event described in section VII of~\cite{Shen2007A} and show that we recover exactly the results obtained with the L-S formalism. As mentioned in the main text, the L-S formalism gives access to the long-time limit of the outgoing wavepackets. We will therefore compare those outgoing wavepackets which have been obtained in the frequency domain with our results which are derived in the time domain.

The specific input state considered in~\cite{Shen2007A} consists of two photons incoming from the left, such that only $\xi_2(\tau_1,\tau_2)$ contributes to the initial state \eref{eq:lightIn}. The outgoing wavepackets in the frequency domain were then found to be
\begin{eqnarray}\label{eq:0SF}
	f^{\cite{Shen2007A}}_0(\omega_1,\omega_2)&=&\lim\limits_{t\to\infty}\bra{\varnothing}\hat{a}^{\leftharpoonup}_{\omega_2}(t)\,\hat{a}^{\leftharpoonup}_{\omega_1}(t)\ket{\psi_\mathrm{in}}/\sqrt{2}\\
	&=&\frac{1}{2}\int\!\rmd \omega_1'\int\!\rmd \omega_2'\,\,\xi_2(\omega_1',\omega_2')\,S^{\leftharpoonup\leftharpoonup,\rightharpoonup\rightharpoonup}_{\omega_1\omega_2,\omega_1'\omega_2'}\nonumber\\
	&=&r_{\omega_1}r_{\omega_2}\,\xi_2(\omega_1,\omega_2)+B^{\cite{Shen2007A}}(\omega_1,\omega_2)\nonumber ,
\end{eqnarray}
\begin{eqnarray}\label{eq:1SF}
	f^{\cite{Shen2007A}}_1(\omega_1,\omega_2)&=&\lim\limits_{t\to\infty}\bra{\varnothing}\hat{a}^{\leftharpoonup}_{\omega_2}(t)\,\hat{a}^{\rightharpoonup}_{\omega_1}(t)\ket{\psi_\mathrm{in}}\\
	&=&\frac{1}{\sqrt{2}}\int\!\rmd \omega_1'\int\!\rmd \omega_2'\,\,\xi_2(\omega_1',\omega_2')\,S^{\rightharpoonup\leftharpoonup,\rightharpoonup\rightharpoonup}_{\omega_1\omega_2,\omega_1'\omega_2'}\nonumber\\
	&=&\sqrt{2}\,\Big(t_{\omega_1}r_{\omega_2}\,\xi_2(\omega_1,\omega_2)+B^{\cite{Shen2007A}}(\omega_1,\omega_2)\Big)\nonumber ,
\end{eqnarray}
\begin{eqnarray}\label{eq:2SF}
	f^{\cite{Shen2007A}}_2(\omega_1,\omega_2)&=&\lim\limits_{t\to\infty}\bra{\varnothing}\hat{a}^{\rightharpoonup}_{\omega_2}(t)\,\hat{a}^{\rightharpoonup}_{\omega_1}(t)\ket{\psi_\mathrm{in}}/\sqrt{2}\\
	&=&\frac{1}{2}\int\!\rmd \omega_1'\int\!\rmd \omega_2'\,\,\xi_2(\omega_1',\omega_2')\,S^{\rightharpoonup\rightharpoonup,\rightharpoonup\rightharpoonup}_{\omega_1\omega_2,\omega_1'\omega_2'}\nonumber\\
	&=&t_{\omega_1}t_{\omega_2}\,\xi_2(\omega_1,\omega_2)+B^{\cite{Shen2007A}}(\omega_1,\omega_2)\nonumber ,
\end{eqnarray}
where we have used (128), (130) and (127) of~\cite{Shen2007A} to substitute the respective scattering matrix elements $S^{\leftharpoonup\leftharpoonup,\rightharpoonup\rightharpoonup}_{\omega_1\omega_2,\omega_1'\omega_2'}$, $S^{\rightharpoonup\leftharpoonup,\rightharpoonup\rightharpoonup}_{\omega_1\omega_2,\omega_1'\omega_2'}$ and $S^{\rightharpoonup\rightharpoonup,\rightharpoonup\rightharpoonup}_{\omega_1\omega_2,\omega_1'\omega_2'}$. Here $t_\omega=1+r_\omega=\omega/(\omega+i)$ is the single-photon transmission amplitude and 
\begin{equation}\label{eq:BSF}
	\!\!\!\!\!\!\!\!\!\!\!\!B^{\cite{Shen2007A}}(\omega_1,\omega_2)= \frac{1}{2\pi}(r_{\omega_1}+r_{\omega_2})\int\!\rmd \omega'\,r_{\omega'}r_{\omega_1+\omega_2-\omega'}\,\xi_2(\omega',\omega_1+\omega_2-\omega')
\end{equation}
is the nonlinear correction that we expect to be equivalent to what we obtained in the time domain~\eref{eq:nonlinCorrection}. As a side remark, the non-trivial form it takes in the frequency domain is already a hint of why the intuitive time domain approach might be more suitable to extend the description to more input photons.

We now proceed with our method, which gives in the time domain
\begin{eqnarray}
	\!\!\!\!\!\!\!\!f_0(\tau_1,\tau_2)&=&\lim\limits_{t\to\infty}\bra{\varnothing}\hat{a}^{\leftharpoonup}_{\tau_2}(t)\,\hat{a}^{\leftharpoonup}_{\tau_1}(t)\ket{\psi_\mathrm{in}}/\sqrt{2}\\
	&=&\frac{1}{\sqrt{2}}\bra{\varnothing}\mathcal{T}\Big[\hat{\sigma}_-(\tau_2)\,\hat{\sigma}_-(\tau_1)\Big]\ket{\psi_\mathrm{in}}\nonumber ,
\end{eqnarray}
\begin{eqnarray}
	\!\!\!\!\!\!\!\!f_1(\tau_1,\tau_2)&=&\lim\limits_{t\to\infty}\bra{\varnothing}\hat{a}^{\leftharpoonup}_{\tau_2}(t)\,\hat{a}^{\rightharpoonup}_{\tau_1}(t)\ket{\psi_\mathrm{in}}\\
	&=&\bra{\varnothing}\hat{\sigma}_-(\tau_2)\,\hat{a}^{\rightharpoonup}_{\tau_1}(0)+\mathcal{T}\Big[\hat{\sigma}_-(\tau_2)\,\hat{\sigma}_-(\tau_1)\Big]\ket{\psi_\mathrm{in}}\nonumber ,
\end{eqnarray}
\begin{eqnarray}
	\!\!\!\!\!\!\!\!f_2(\tau_1,\tau_2)&=&\lim\limits_{t\to\infty}\bra{\varnothing}\hat{a}^{\rightharpoonup}_{\tau_2}(t)\,\hat{a}^{\rightharpoonup}_{\tau_1}(t)\ket{\psi_\mathrm{in}}/\sqrt{2}\\
	&=&\frac{1}{\sqrt{2}}\Big(\bra{\varnothing}\hat{a}^{\rightharpoonup}_{\tau_2}(0)\,\hat{a}^{\rightharpoonup}_{\tau_1}(0)+\hat{\sigma}_-(\tau_1)\,\hat{a}^{\rightharpoonup}_{\tau_2}(0)+\hat{\sigma}_-(\tau_2)\,\hat{a}^{\rightharpoonup}_{\tau_1}(0)+\mathcal{T}\Big[\hat{\sigma}_-(\tau_2)\,\hat{\sigma}_-(\tau_1)\Big]\ket{\psi_\mathrm{in}}\Big)\nonumber .
\end{eqnarray}
Would our goal be to solve the scattering event, we could stop here and use our main result to express the nonlinear term as presented in the main text. However in order to compare with~\cite{Shen2007A}, we will now inverse Fourier transform the wavepackets into the frequency domain. As far as the linear part is concerned, this is a trivial step with the use of the convolution theorem on \eref{eq:triviallyLin} and \eref{eq:ifLinearBS}
\begin{eqnarray}
	\mathcal{F}^{-1}\Big[\bra{\varnothing}\hat{a}^{\rightharpoonup}_{\tau_2}(0)\,\hat{a}^{\rightharpoonup}_{\tau_1}(0)\ket{\psi_\mathrm{in}}\Big](\omega_1,\omega_2)= \sqrt{2}\,\xi_2(\omega_1,\omega_2) ,\\
	\mathcal{F}^{-1}\Big[\bra{\varnothing}\hat{\sigma}^\mathrm{lin}_-(\tau_2)\,\hat{a}^{\rightharpoonup}_{\tau_1}(0)\ket{\psi_\mathrm{in}}\Big](\omega_1,\omega_2)= \sqrt{2}\,r_{\omega_2}\,\xi_2(\omega_1,\omega_2) \nonumber ,\\
	\mathcal{F}^{-1}\Big[\bra{\varnothing}\hat{\sigma}^\mathrm{lin}_-(\tau_2)\,\hat{\sigma}^\mathrm{lin}_-(\tau_1)\ket{\psi_\mathrm{in}}\Big](\omega_1,\omega_2)=\sqrt{2}\,r_{\omega_1}r_{\omega_2}\,\xi_2(\omega_1,\omega_2)\nonumber ,
\end{eqnarray}
and we straightforwardly recover the linear contributions in \eref{eq:0SF}, \eref{eq:1SF} and \eref{eq:2SF}. We are thus left with the nonlinear correction~\eref{eq:nonlinCorrection}
\begin{eqnarray}
	B(\tau_1,\tau_2)&=&\frac{1}{\sqrt{2}}\Big(\bra{\varnothing}\mathcal{T}\Big[\hat{\sigma}_-(\tau_2)\,\hat{\sigma}_-(\tau_1)\Big]-\hat{\sigma}^\mathrm{lin}_-(\tau_2)\,\hat{\sigma}^\mathrm{lin}_-(\tau_1)\ket{\psi_\mathrm{in}}\Big)\\
	&=&-\theta(\tau_2-\tau_1)\rme^{-(\tau_2-\tau_1)}\int_{0}^{\tau_1}\!\rmd t''\,\int_{0}^{\tau_1}\!\rmd t'\,\rme^{-(\tau_1-t'')}\rme^{-(\tau_1-t')}\xi_2(t',t'')+(\tau_1\leftrightarrow\tau_2)\nonumber .
\end{eqnarray}
The key to proceed further is to inverse Fourier transform along the $\tau_2$ variable first ($\tau_1$ for the second term) which yields $r_{\omega_2}\rme^{i\omega_2\tau_1}/\sqrt{2\pi}$ ($r_{\omega_1}\rme^{i\omega_1\tau_2}/\sqrt{2\pi}$ for the second term). We can then use the following property which is valid for any two-dimensional function $g(\omega_1,\omega_2)$ with a well-defined Fourier transform $\mathcal{F}[g(\omega_1,\omega_2)](\tau_1,\tau_2)=g(\tau_1,\tau_2)$
\begin{eqnarray}
	&\mathcal{F}^{-1}&[g(\tau,\tau)](\omega)=\frac{1}{\sqrt{2\pi}}\int\rmd \omega'\,g(\omega',\omega-\omega') \\
	&\Longrightarrow&\ \mathcal{F}^{-1}\Big[\int_{0}^{\tau}\!\rmd t''\,\int_{0}^{\tau}\!\rmd t'\,\rme^{-(\tau-t'')}\rme^{-(\tau-t')}\xi_2(t',t'')\Big](\omega_1+\omega_2)\nonumber \\
	&&=\frac{1}{\sqrt{2\pi}}\int\rmd \omega'\,r_{\omega'}r_{\omega_1+\omega_2-\omega'}\,\xi_2(\omega',\omega_1+\omega_2-\omega') \nonumber,
\end{eqnarray}
such that we recover exactly the nonlinear term~\eref{eq:BSF}, which concludes our comparison with~\cite{Shen2007A}. 

\section*{References}
\bibliographystyle{iopart-num.bst}
\bibliography{Bibliography}

\end{document}